\documentclass[a4paper]{jpconf}
\usepackage{amsmath}
\usepackage{hyperref}
\usepackage{graphicx}
\usepackage{float}
\begin{document}
\title{LHCb trigger streams optimization}
\author{
  D Derkach\textsuperscript{1, 2}, N Kazeev\textsuperscript{1, 2},
  R Neychev\textsuperscript{2, 3}, A Panin\textsuperscript{2},
  I Trofimov\textsuperscript{4}, A Ustyuzhanin\textsuperscript{1, 2, 3}
  and M Vesterinen\textsuperscript{5}}
\address{\textsuperscript{1} National Research University Higher School of Economics (HSE),
  Moscow, Russia}
\address{\textsuperscript{2} Yandex~School~of~Data~Analysis (YSDA), Moscow, Russia}
\address{\textsuperscript{3} Moscow~Institute~of~Physics~and~Technology (MIPT), Moscow, Russia}
\address{\textsuperscript{4} Yandex~Data~Factory (YDF), Moscow, Russia}
\address{\textsuperscript{5} Ruprecht-Karls-Universitaet~Physikalisches~Institut, Heidelberg, Germany}
\ead{nikita.kazeev@cern.ch}

\begin{abstract}
  The LHCb experiment stores around $10^{11}$ collision events per
  year. A typical physics analysis deals with a final sample of up to
  $10^7$ events. Event preselection algorithms (lines) are used for
  data reduction. Since the data are stored in a format that requires
  sequential access, the lines are grouped into several output file
  streams, in order to increase the efficiency of user analysis jobs
  that read these data. The scheme efficiency heavily depends on the
  stream composition. By putting similar lines together and balancing
  the stream sizes it is possible to reduce the overhead. We present a
  method for finding an optimal stream composition. The method is
  applied to a part of the LHCb data (Turbo stream) on the stage where
  it is prepared for user physics analysis. This results in an
  expected improvement of 15\% in the speed of user analysis jobs, and
  will be applied on data to be recorded in 2017.
\end{abstract}

\section{Introduction}
To capture and analyze a large number of collision events, the LHCb
experiment \cite{LHCb:2003ab} relies on a multi stage data processing
pipeline \cite{Aaij:2012me}. The events are filtered through the
hardware L0 trigger and two levels of software triggers – HLT1 and
HLT2. Physicists develop algorithms (called lines) that select
particular types of events that they wish to study. All the events
that satisfy the requirements of at least one HLT2 selection line are
permanently recorded to tape storage. Since Run-II, the HLT2 output
data have been split into two streams. Data in the FULL stream need to
be reconstructed on distributed computing resources and are intended
for further event selection, before being made available for user
analysis. Run-II saw the introduction of the Turbo stream, with a
event format ready for analysis right after the trigger step, without
further event preselection. Turbo stream data are prepared for physics
analysis by an application called Tesla~\cite{aaij2016tesla}.

User analysis jobs run independently and usually require only a small
subset of all events selected by the lines. Efficient data storage and
access methods are therefore required. LHCb uses the Worldwide LHC
Computing Grid (WLCG), which supports data granularity on file
level~\cite{wlcg-man}.

The LHCb experiment uses two factors to group events into
files. First, a file only contains events from a single run. In LHCb a
run corresponds to a period of up to 1 hour of data taking, in which
the beam and detector conditions are presumed constant. This makes it
easier to apply different collision conditions and discard runs that
are flagged up by the data quality assessment. Second, the lines are
grouped into streams, such that each file available for user analysis
corresponds to a particular run-stream pair. If an event passes lines
from different streams, it (wholly or partially) will be copied to
multiple files. Sets of files corresponding to particular streams are
themselves also called streams~\cite{lhcb-computing-model}. Both FULL
and Turbo stream are further divided into streams. To avoid confusion
in this paper we refer to the streams into which the Turbo stream is
divided as Tesla streams.

\section{Optimization criteria}
Several considerations are made when defining the mapping of lines to
streams.
\begin{itemize}
\item User job performance. A job has to read a whole Tesla stream
  even if it needs only a small subset of events in it. The optimum is
  achieved if each line is assigned to a separate stream. For Tesla
  streams, the estimated time spent by the user jobs on disk access
  differs by a factor of 5 between the extreme variants. The metric is
  described in Sect.~\ref{sec:T}~and~\ref{sec:Treal}.
\item Storage space. Information is duplicated when an event belongs
  to multiple streams. The optimal storage performance would be
  achieved if all lines belong to a single stream. For Tesla streams,
  the scheme where each line is assigned to a separate stream will
  take 1.5x more space than a single stream. The evaluation procedure
  for storage space usage is described in Sect.~\ref{sec:Validation}.
\end{itemize}

These factors must be estimated in order to construct a streaming
scheme. There is another constraint — WLCG often uses tape storage
systems, which generally do not cope well with storing and providing
frequent access to many small files \cite{lhcb-distributed-da}. Since
each stream has at least one file for each run, the number of files
grows with the number of streams. The other problem is the management
of those streams for the data management team. More streams need more
operations on them (replication, deletion, also staging).

\subsection{Disk access time}
\label{sec:T}
The total time spent by user jobs on disk access depends on two
independent factors: the queries the users make and the time it takes
to complete each query. We use the following assumptions:
\begin{itemize}
\item The number of times each event is requested is proportional to
  the number of lines it passes.
\item The time that a job would spend on disk access is proportional
  to the number of events in the stream that the job reads.
\end{itemize}

So the total time would be proportional to
\begin{equation}
  \label{eq:T}
  T = \sum_\text{streams} N_\text{events in stream} \cdot N_\text{lines in stream}
\end{equation}

Some lines are prescaled – line positive selection decisions are
randomly discarded with a specified probability. This can be
accommodated by using the expected value of $T$:
\begin{equation}
  \label{eq:T:prescaled}
  E\left(T\right) = \sum_\text{streams} N_\text{lines in stream} \cdot
  E\left(N_\text{events in stream}\right),
\end{equation}
\begin{equation}
  E[N_\text{events in stream}] = \sum_e \left(1 - \prod_l\left(1 - \Delta_{el}P_lL_{ls}\right)\right),
\end{equation}
where $\Delta_{el} \in \{0, 1\}$ is the indicator whether event $e$
passes line $l$, $P_l \in [0, 1]$ is line $l$ prescale value,
$L_{ls} \in \{0, 1\}$ is the indicator whether line $l$ belongs to
stream $s$.

\subsection{Disk usage}
\label{sec:S}
The amount of information about an event recorded to a particular
stream has a non-trivial relationship with the set of lines from that
stream that have selected the event. We use a simplified model in
which there are 2 types of Turbo lines. Pure Turbo lines store the
information about the decay candidate that triggers the selection of
the event. These are assumed to have a size of $10$ kB per line. In
2016, the Turbo stream was extended to allow full event reconstruction
information to be persisted. These lines with the PersistReco flag are
assumed to store an additional $50$ kB shared among such
lines. Thus, the formula for event $e$ size $S_{es}$ in the stream
$s$:
\begin{equation}
  S_{es} = 10 \cdot N_\text{turbo lines} + 50 \cdot I_\text{persist reco}
\end{equation}
where $N_\text{turbo lines} \in \mathbf{N}_0$ is the number of lines
with the Turbo flag belonging to the stream $s$ that the event passes,
$I_\text{persist reco} \in \{0, 1\}$ is the indicator of whether the
event passes a line with the PersistReco flag belonging to the stream.

\section{Optimization}
This is a clustering problem – we want lines that select similar
events to be grouped together. We have tried classic clustering
algorithms from scikit-learn \cite{pedregosa2011scikit}: KMeans,
SpectralClustering, Birch, AffinityPropagation. They have failed to
improve the baseline, which is not surprising given that the loss
functions used are quite different from ours. We therefore postulate
that our algorithm must
\begin{itemize}
\item optimize $T$ directly instead of some cluster goodness function
\item allow for different cost functions to be able to utilize a more
  accurate model in the future
\item converge to a reasonable solution within a reasonable time
\item accept the number of streams as a parameter to maintain the WLCG
  constraint on the number of files
\end{itemize}

\subsection{Continuous loss}
The first step is the transition from discrete to continuous
optimization to be able to use fast gradient methods. Instead of
assigning the lines to streams, we let each line $l$ have a
probability $\tilde{L}_{ls}$ to be in each stream $s$.
\begin{equation}
  \label{eq:EL}
  E\left[N_\text{lines in stream}\right] = \sum_l \tilde{L}_{ls},
\end{equation}
\begin{equation}
  \label{eq:EE}
  E\left[N_{\text{events in stream }s}\right] = \sum_e\left(1 - \prod_l\left(
      1 - \Delta_{el}P_l\tilde{L}_{ls}\right)\right).
\end{equation}
After optimization, we assign each line to the stream with the highest
probability of containing it.

\begin{eqnarray}
  \tilde{T} = \sum_s E\left[N_{\text{lines in stream }s}\right] \cdot
  E\left[N_{\text{events in stream }s}\right]\\
  = \sum_s\left[ \sum_l \tilde{L}_{ls} \cdot \sum_e\left(1 - \prod_l\left(
  1 - \Delta_{el}P_l\tilde{L}_{ls}\right)\right) \right].
\end{eqnarray}
In general, $\tilde{T} \neq E[T]$. However, if all the assignments are
definite, $L_{ls}\in\{0,1\}$ and $\tilde{T} = T$. In practice on our
data, the algorithm has nearly always converged to near integer
probabilities.

\subsection{Solving the boundary conditions}
Since $\tilde{L}_{ls}$ are probabilities, there are constraints on
their values:
\begin{itemize}
\item $\tilde{L}_{ls}\in[0,1]$
\item the sum of the probabilities of all random outcomes must be 1,
  $\sum_s \tilde{L}_{ls} = 1$
\end{itemize}
To satisfy these conditions, we have parametrized $\tilde{L}_{ls}$ with
softmax \cite{bishop2006pattern}:
\begin{equation}
  \label{eq:softmax}
  \tilde{L}_{ls} = \frac{e^{A_{ls}}}{\sum_s e^{A_{ls}}},
\end{equation}
$A_{ls}$ can have any value.

\subsection{Grouping constraints}
Line usage is not independent. Some lines are often requested
together. Therefore, from the point of view of user convenience, it is
desirable to have those lines in a single stream. From now on, such
groups will be referred to as modules.

The formula for $\tilde{T}$ can be parametrized to make the result strictly
adhere to the grouping requirement:
\begin{equation}
  \tilde{\Delta}_{em} = 1 - \prod_l\left(1 - \Delta_{el}P_lM_{lm}\right),
\end{equation}
\begin{equation}
  \label{eq:Tmodules}
  \tilde{\tilde{T}} = \sum_s\left[ \left( \sum_m \sum_l M_{lm} \tilde{L}_{ms}\right) \cdot
    \sum_e\left(1 - \prod_m\left(1 - \tilde{\Delta}_{em}\tilde{L}_{ms}\right)\right) \right],
\end{equation}
where $M_{lm}\in \{0, 1\}$ is the indicator of whether the module $m$
contains the line $l$, $\tilde{L}_{ms}$ is the probability of module
$m$ being in stream $s$, $\tilde{\Delta}_{em}$ is the probability that
event $e$ was selected by the module $m$.

\subsection{Implementation}
The optimization is implemented in Python using the theano framework
\cite{team2016theano}. Theano is a mature framework primarily used for
deep learning. Its major advantages for our task are speed and
symbolic gradients computation. We have tested several gradient
optimization algorithms: Nesterov Momentum, AdaMax, AdaGrad, AdaM,
AdaDelta, AdaMax. For our data the best results are achieved with
AdaMax.

The code is freely available under Apache License 2.0:
\url{https://gitlab.cern.ch/YSDA/streams-optimization/}

\section{Results for the LHCb Turbo stream}
The method is applied to find an optimal Tesla stream composition for
the LHCb Turbo stream. We take a sample of $10^{5}$ Run-II Turbo
events recorded in October 2016 and compare the optimized streams to
the baseline where the lines are grouped by physical similarity
\cite{baseline}.

HLT lines are grouped into separate modules that are typically
authored by a small team. They tend to contain several selections that
are topologically similar and/or are required for a single analysis or
set of related analyses. One of the modules contains several hundred
selections of charm hadron decay selections. This module is divided
into submodules. For user convenience, we do not split the modules and
charm hadron submodules.

\subsection{Model validation}
\label{sec:Validation}
We stream the events with various streaming schemes – produced by our
algorithm for different stream numbers and the baseline. We measure
the files sizes and the time needed to process the files with a
minimalist analysis job using the DaVinci application
\cite{lhcb2015davinci}, which only reads events and lists HLT decision
flags in them. For each stream in each scheme, we calculate the $T$
and $S$ values and calibrate them by fitting a least squares linear
regression. For the calibrated values, we compute the coefficient of
determination. The results are as follows:
\begin{equation}
  R^2(T) = 0.57,
\end{equation}
\begin{equation}
  R^2(S) = 0.98.
\end{equation}

The model is good in describing event sizes. For reading times, the
model is not perfect. It appears that the time it takes to read an
event depends on the event structure.

Job run time on a real world computer is subject to random
fluctuations. To estimate their influence, we rerun the DaVinci job 5
times. The standard deviation of the results is 2\%.

\subsection{Results}
\label{sec:Treal}
\begin{figure}[H]
  \centering
  \includegraphics[width=\textwidth]{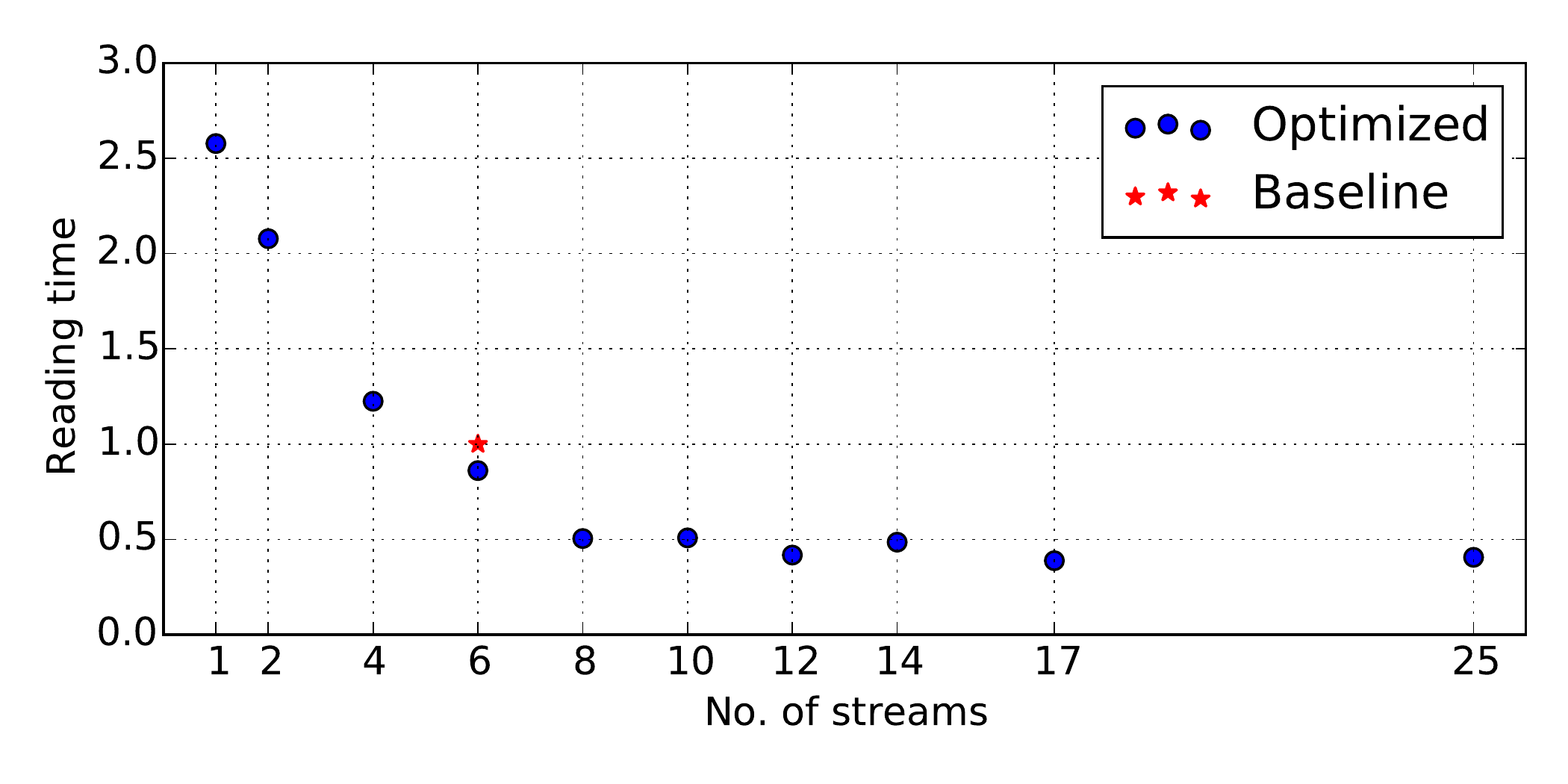}
  \caption{Evaluation of optimized streams. On the vertical axis
    $T_\text{real}$ is plotted with values normalized to the baseline,
    where the lines are grouped by physical similarity
    \cite{baseline}. For the same number of streams, the disk reading
    time of the analysis jobs is improved by 15\%, while adding two
    additional streams brings this to 50\%.}
  \label{fig:T-n-streams}
\end{figure}

We build optimized schemes for different numbers of streams. Then we
stream the events and, for each resulting file, measure size and
reading time $T_\text{stream}$. The difference in total file sizes is
less than 2\%. For the reading times, we apply the same query
assumptions we used in our model – the number of times each event is
requested is proportional to the number of lines it passes. The test
job also takes some time to initialize –
$T_\text{initial} \approx 9\ \text{s}$. This is a function of the job
and not of the streaming scheme, and is accounted for.
\begin{equation}
  T_\text{real} = \sum_\text{streams} N_\text{lines in stream} \left(
    T_\text{stream} - T_\text{initial}\right).
\end{equation}
The results are presented in Figure \ref{fig:T-n-streams}.

\section{Conclusion}
We present a method for finding the optimal stream composition. It is
flexible and can be used for different cost functions and numbers of
streams. For the Tesla streams, it is possible to decrease the disk
reading time of the analysis jobs by 15\% while maintaining the line
groupings and stream counts.

\section*{References}
\bibliography{research}
\end{document}